\documentclass[12pt]{iopart}
\usepackage{times}
\usepackage{graphicx}
\usepackage{amssymb}
\usepackage{lineno}
\usepackage{xcolor}
\usepackage{url,hyperref}
\newcommand{\resampp}{\tiny\mbox{resamp}}
\newcommand{\finall}{\mbox{final}}
\newcommand{\maxx}{\tiny{\mbox{max}}}
\newcommand{\modell}{\tiny{\mbox{model}}}

\newcommand{\RR}{\tiny{\mbox{R}}}
\newcommand{\KK}{\tiny{\mbox{K}}}
\newcommand{\rann}{\tiny{\mbox{ran}}}
\newcommand{\tott}{\tiny{\mbox{tot}}}
\newcommand{\QQ}{\tiny{\mbox{Q}}}
\newcommand{\obss}{\tiny{\mbox{obs}}}
\newcommand{\calcc}{\tiny{\mbox{calc}}}
\newcommand{\corr}{\tiny{\mbox{cor}}}
\newcommand{\WW}{\tiny{\mbox{W}}}
\newcommand{\JJ}{\tiny{\mbox{J}}}
\newcommand{\sss}{\tiny{\mbox{s}}}
\begin{document}

\title[
Spectral model
selection 
in the electronic measurement of the Boltzmann constant
by Johnson noise thermometry
]{Spectral model selection
in the electronic measurement of the Boltzmann constant
by Johnson noise thermometry
}

\author{Kevin J Coakley$^1$, Jifeng Qu$^2$}

\address{$^1$ National Institute of Standards and Technology (NIST),
Boulder, CO 80302 USA}
\address{$^2$
National Institute of Metrology (NIM), Beijing 100029, People’s Republic of China}
\ead{kevin.coakley@nist.gov}

\section{Abstract}
In the electronic measurement of the Boltzmann constant based on
Johnson noise thermometry, the ratio of the power spectral densities
of thermal noise across a resistor at the triple point of water,
and pseudo-random noise synthetically
generated by a quantum-accurate voltage-noise source
is constant to within 1 part in a billion for frequencies up to 1 GHz.
Given knowledge of this ratio, and the values of other parameters that are known or measured,
one can determine the Boltzmann constant.
Due, in part, to
mismatch between transmission lines,
the experimental ratio spectrum
varies with frequency.
We model this spectrum 
as an even polynomial function of frequency
where the constant term 
in the polynomial
determines the Boltzmann constant.
When determining this constant (offset) from experimental data,
the assumed complexity of the ratio spectrum
model and the maximum frequency analyzed (fitting bandwidth)
dramatically affects results.
Here, we
select
the complexity of the model
by cross-validation -- a data-driven statistical learning method.
For each of many fitting bandwidths,
we determine the component of uncertainty
of the offset term
that accounts for random
and systematic effects associated with imperfect knowledge of model
complexity.
We select the fitting bandwidth that minimizes
this uncertainty.
In the most recent measurement of the Boltzmann constant,
results were determined, in part, by application of an earlier version of
the
method described here.
Here, we extend the earlier analysis by considering a broader
range of fitting bandwidths and quantify an additional component of
uncertainty that accounts for imperfect
performance of our fitting bandwidth selection method.
For idealized simulated data
with additive noise
similar  
to  experimental data,
our method correctly selects the true complexity
of the ratio spectrum model for all
cases considered.
A new analysis of 
data
from the recent experiment yields
evidence for a temporal trend
in the offset parameters.
\\
Keywords:
Boltzmann constant,
cross-validation,
Johnson noise thermometry, 
model selection,
resampling methods,
impedance mismatch

\section{Introduction}

There are various experimental methods to determine the Boltzmann
constant
including
acoustic gas 
thermometry \cite{M1},\cite{P1},\cite{Pod1}; dielectric constant gas thermometry \cite{G1},\cite{L1}, \cite{Ga1}, \cite{Ga2}, Johnson noise thermometry (JNT)
\cite{B1}, \cite{B2},
\cite{Qu01}, and Doppler broadening \cite{Bo1}, \cite{F1}.
CODATA
(Committee on Data for Science and Technology)
will
determine the
Boltzmann constant as a weighted average of estimates determined
with these methods.
Here, we focus on JNT experiments which utilize 
a quantum-accurate voltage-noise source (QVNS).

In JNT, one infers true thermodynamic
temperature based on measurements of the fluctuating voltage and
current noise caused by the random thermal motion of electrons in
all conductors. According to the Nyquist law, the mean square of
the fluctuating voltage noise for frequencies below 1 GHz and
temperature near 300 K can be approximated to better than 1 part
in  billion as $<V^2> = 4 kT R \Delta f$, where $k$ is the Boltzmann
constant, $T$ is the thermodynamic temperature, $R$ is the resistance
of the sensor, and $\Delta f$  is the bandwidth over which the noise
is measured.
Since JNT is a pure electronic
measurement
that is
immune from chemical and mechanical changes in
the material properties of the sensor,
it is an appealing alternative to other forms of
primary gas thermometry that are limited by the non-ideal properties
of real gases.

Recently, interest in JNT has dramatically increased because of its
potential contribution to the ``New SI" (New International System),
in which the unit of thermodynamic temperature, the kelvin, will
be redefined in 2018 by fixing the numerical value of $k$. Although
almost certainly the value of $k$ will be primarily determined by the
values obtained by acoustic gas thermometry, there remains the
possibility of unknown systematic effects that might bias the
results, and therefore an alternative determination using a different
physical technique and different principles is necessary to ensure
that any systematic effects must be small. To redefine the kelvin,
the Consultative Committee on Thermometry (CCT) of the International
Committee for Weights and Measures (CIPM) has required that besides
the acoustic gas thermometry method, there must be another method
that can determine $k$ with a relative uncertainty below 3$\times
10^{-6}$. As of now, JNT is the most likely method to meet this
requirement.

\textcolor{black}{
In JNT with a QVNS,
\cite{Qu01},
according to physical theory,
the Boltzmann
constant is related to the ratio of the power spectral density \textcolor{black}{(PSD)}
of noise produced by
a resistor at the
triple point of water temperature
and
the \textcolor{black}{PSD} of
noise produced
by
a QVNS.
For, frequencies below 1 GHz, this ratio
is constant to within 1 part in $10^{9}$.  
The physical model for the
\textcolor{black}{PSD}
for the noise across the resistor is
$S_{\RR}$
where
\begin{eqnarray}
S_{\RR} = 4 k T_{\WW} X_{\RR} R,
\end{eqnarray}
$T_{\WW}$ is the temperature of the triple point of water,
$X_{\RR}$ is the resistance in units of the von Klitzing resistance $R_{\KK} = \frac{h}{e^2}$
where $e$ is the charge of the electron and $h$ is Planck's constant.
The model for the
PSD of the
noise produced by the QVNS,
$S_{\QQ}$, 
is
\begin{eqnarray}
S_{\QQ} = D^2 N^2_{\JJ} f_{\sss} M / K^2_{\JJ},
\end{eqnarray}
where $K_{\JJ} = \frac{2e}{h}$, $f_{\sss}$ is a clock frequency, $M$ is a bit length
parameter, $D$ is a software input parameter that determines the 
amplitude of the QVNS waveform, and $N_{\JJ}$ is the number of junctions
in the Josephson array 
in the QVNS.
\textcolor{black}{Assuming that the  Eq. 1 and Eq. 2 models
are valid,
the Boltzmann constant $k$ is}
\begin{eqnarray}
k = h \frac{ D N^2_{\JJ} f_{\sss} M}
{ 16 T_{\WW}
X_{\RR} }
\frac{S_{\RR}}
{S_{\QQ}}.
\end{eqnarray}
}

In actual experiments,
transmission lines
that
connect the resistor and the QVNS to preamplifiers
produce a
ratio spectrum that varies with frequency.
Due solely to
impedance mismatch effects, 
for the frequencies of interest,
one expects the 
ratio spectrum \textcolor{black}{predicted by physical theory}
to be an
even polynomial function of frequency where the 
constant term (offset) in the polynomial is the
value $S_{\RR}/S_{\QQ}$
provided that dielectric losses are negligible.
The theoretical justification for this polynomial model
is based on low-frequency filter theory where
measurements are modeled
by a ``lumped-parameter approximation."
In particular,
one models the networks
for the resistor and the QVNS
as combination of series and parallel complex impedances
where
the impedance coupling in the resistor network
is somewhat different from that in the QVNS network.
For the ideal case where
all shunt capacitive
impedances are real, there are no dielectric losses.
As a caveat,
in actual experiments,
other effects including electromagnetic
interference
and filtering aliasing also affect
the ratio spectrum.
As discussed in \cite{Qu01},
for the the recent experiment of interest,
dielectric losses and other effects 
are small compared to impedance mismatch effects.
As an aside,
impedance mismatch 
effects also influence
results in
JNT experiments 
that do not utilize a QVNS 
\cite{Kamp}, \cite{Bla}, \cite{Whi}.

Based on a fit of the ratio spectrum model to the observed ratio spectrum,
one determines the offset parameter $S_{\RR}/S_{\QQ}$.
Given this estimate of $S_{\RR}/S_{\QQ}$ and values of other terms 
on the right hand sided of Eq. 3 (which are known or measured),
one determines the Boltzmann constant.
However,
the choice
of the order $d$ (complexity) of the polynomial ratio spectrum  model and the upper frequency cutoff for analysis
(fitting bandwidth $f_{\maxx}$) 
significantly affects both the estimate and its associated uncertainty.
In JNT, researchers typically select the model complexity and fitting bandwidth based
on scientific judgement informed by graphical analysis of
results.
A common approach is to restrict attention to sufficiently
low fitting bandwidths where curvature in the ratio spectrum is
not too dramatic and
assume that a quadratic spectrum model
is valid
(see \cite{B2} for an example of this approach).

In contrast to a practitioner-dependent subjective approach,
we present a
data-driven
objective
method to select the complexity of the
ratio spectrum model and the fitting bandwidth.
In particular, we select the ratio spectrum model based on
cross-validation 
\cite{St1},\cite{St2},\cite{AC},\cite{HTF}.
We note that in addition to cross-validation,
there are other model selection methods
including
those
based on the Akaike Information Criterion (AIC) 
\cite{Ak}, the Bayesian Information Criterion (BIC) \cite{S1},
$C_p$ statistics \cite{Mallows}.
However,
cross-validation 
is more data-driven and flexible than these other 
approaches because
it relies on
fewer modeling assumptions.
Since the selected model
determined by any method is a function of 
random data, none perform perfectly.
Hence,
uncertainty due to imperfect
model
selection performance
should be accounted for 
in the 
uncertainty budget for any parameter 
of interest
\cite{BA}, \cite{CH}.
Failure to account for uncertainty in the selected model
generally leads to underestimation of uncertainty.

In cross-validation, one
splits observed data into
training and validation subsets.
One 
fits candidate models to training data,
and selects the model that is most consistent
with validation data that are independent of the training
data. 
Here (and in most studies) consistency is measured
with a cross-validation statistic equal to
the mean square deviation between predicted and observed values
in validation data.
We stress that, in general,
this mean square deviation depends on both random and systematic
effects.

For each candidate model,
practitioners sometimes
average cross-validation statistics from many
random splits of the data
into training and validation data set
\cite{Pic},
\cite{Sh}, \cite{Xu}.
Here,
from many random splits,
we instead determine
model selection fractions
determined from a five-fold cross-validation
analysis.
Based on these model selection fractions,
we determine
the uncertainty
of 
the offset parameter for each fitting bandwidth of interest.
We select
$f_{\maxx}$ by minimizing
this uncertainty.
As far as we know, our
resampling
approach for quantification of uncertainty due to random effects and 
imperfect performance of
model selected by
five-fold
cross-validation 
is new.
As an aside,
for the case where
models are selected based on
$C_p$
statistics,
Efron \cite{Ef1}
determined
model selection
fractions with a 
bootstrap resampling scheme.


In \cite{Qu01}, 
the complexity of the ratio spectrum
and the fitting bandwidth
were selected
with an earlier version of the method described 
here
for the case where 
$f_{\maxx}$ was no greater than 600 kHz.
Here, we re-analyze
the data from \cite{Qu01}
but allow
$f_{\maxx}$ to be as large as
1400 kHz.
In this work, we also
quantify an additional component of uncertainty
that accounts for
\textcolor{black}{imperfect performance of our
method for selecting the
fitting bandwidth.}
We stress that this work focuses only
on the uncertainty
of
the offset parameter in ratio spectrum model.
For a discussion of other sources of uncertainty
that affect
the estimate of the Boltzmann constant,
see
$\cite{Qu01}$.
In a simulation study, we show that
our methods 
correctly selects
the correct ratio spectrum for
simulated data with additive noise similar to observed data.
Finally,
for experimental data,
we quantify evidence for a possible 
linear temporal trend in estimates 
for the offset parameter.

\section{Methods}

\subsection{Physical model}
Following \cite{Qu01},
to account for impedance mismatch effects,
we 
model the ratio of the
power spectral densities
\textcolor{black}{of resistor noise and QVNS noise},
${r}_{\textcolor{black}{\modell}}(f)$,
as
a $d$th order even polynomial function of frequency
as follows
\begin{eqnarray}
\textcolor{black}{
{r}_{\textcolor{black}{\modell}}(f) = \sum_{i=0}^{i_{\maxx}} \alpha_{2j} (\frac{f}{f_0})^{2i},
}
\end{eqnarray}
where
$d = 2 i_{\maxx}$,
and
$f_0$ is a reference frequency (1 MHz in this work).
Throughout this work, as shorthand, we
refer to this model as a $d = 2$ model if $i_{\maxx}=1$,
a $d = 4$ model if $i_{\maxx}=2$, and so on.
The constant term, $a_0$ in the Eq. 4 model
corresponds to
${S_{\RR}}/{S_{\QQ}}$
\textcolor{black}{where $S_{\RR}$ and $S_{\QQ}$ are predicted by Eq. 1 and Eq 2. respectively.}

\subsection{Experimental data}
\textcolor{black}{
Data was acquired for each of 45 runs
of the experiment
\cite{Qu01}.
Each run occurred on a distinct day
between
June 12, 2014 to September 10, 2014.}
The time to acquire data \textcolor{black} {for each run} varied from 15 h to 20 h. 
For each run,
Fourier transforms of
time series corresponding to the
resistor at the
triple point of water temperature and
the QVNS
were determined at a resolution of 1 Hz
for frequencies up to
2 MHz.
\textcolor{black}{Estimates of
mean PSD}
were formed for frequency
blocks of width 1.8 kHz.
For the frequency block with midpoint $f$,
for the $i$th run,
we denote
the \textcolor{black}{mean PSD} estimate
for the resistor \textcolor{black}{noise} and QVNS 
\textcolor{black}{noise}
for the $i$th run as
${S_{\RR,\obss}}(f,i)$ 
and
${S_{\QQ,\obss}}(f,i)$ respectively
where $i=$ 1,2, $\cdots$ 45.

Following \cite{Qu01},
we define a reference value
$a_{0,\calcc}$
for the
offset term in our Eq. 4
model as
\begin{eqnarray}
a_{0,\calcc} = \frac{4 k_{2010} RT }{ S_{\QQ,{\calcc}}},
\end{eqnarray}
where $k_{2010}$ is the CODATA2010 recommended value of the
Boltzmann constant, $R$ is the measured resistance of the 
resistor with traceability to the quantum Hall resistance,
$T$ is the triple point water temperature and $S_{\QQ,\calcc}$ is the 
calculated power spectral 
density of \textcolor{black}{QVNS noise}.

\textcolor{black}
{In the recent experiment 
the true value of the resistance, $R$,
could have varied from run-to-run.
Hence,
in \cite{Qu01},
$R$ was measured for each run in a 
calibration experiment.
Based on these calibration experiments,
$a_{0,\calcc}$ was determined for each run.}
The difference between the maximum and minimum of the estimates 
determined from all 45 runs is
2.36 $\times 10^{-6}$.
For the $i$th run, we denote
the
values $a_0$ and $a_{0,\calcc}$ as $a_0(i)$ and
$a_{0,\calcc}(i)$ respectively.
Even though
$a_0(i)$ and 
$a_{0,\calcc}(i)$ 
vary from run-to-run,
we assume that temporal
variation of their difference,
$a_0(i) - 
a_{0,\calcc}(i)$, 
is negligible. 
Later in this work, we check the validity of this 
key modeling assumption.
The component of uncertainty
of the estimate of
$a_{0,\calcc}$ for any run 
due to imperfect knowledge of
$R$
is approximately 2 $\times 10 ^{-7}$.
The estimated weighted mean value of our estimates, 
$\hat{\bar{a}}_{0,\calcc}$,
is 
1.000100961. The weights are determined from 
relative data acquisition times for the runs. 

Following \cite{Qu01},
for each frequency,
we pool data
from all 45 runs to form
a numerator term
$\sum_{i=1}^{45} S_{\RR,\obss}(f,i)$
and a denominator term
$\sum_{i=1}^{45} S_{\QQ,\obss}(f,i)$. 
From these two terms,
we estimate
one ratio 
for each frequency
as
\textcolor{black}{
\begin{eqnarray}
r_{\obss}(f) = 
\frac{ 
\sum_{i=1}^{45} S_{\RR,\obss}(f,i)
}
{
\sum_{i=1}^{45} S_{\QQ,\obss}(f,i) 
}
\end{eqnarray}
}
(see Figure 1).
From the Eq. 6 ratio spectrum,
we estimate one residual offset
term
$a_0 -
{\bar{a}}_{0,\calcc}$
where
${\bar{a}}_{0,\calcc}$
is the weighted mean of $a_{0,\calcc}$
values from all runs.

\subsection{Model selection method}
In our cross-validation approach, we select the model that produces
the prediction (determined from the training data) that is most consistent
with the validation data. 
Since $a_{0,\calcc}$ varies from run-to-run,
we correct $S_{\RR,\obss}$ spectra so that our cross-validation
statistic
is not artificially inflated by
\textcolor{black}{run-to-run 
variations} in $a_{0,\calcc}$.
The corrected ratio spectrum
for the $i$th run is
\textcolor{black}{
\begin{eqnarray}
\textcolor{black}{
{r}_{\corr}(f)} = 
\frac{ 
\sum_{i=1}^{45} S_{\RR,\corr}(f,i)
}
{
\sum_{i=1}^{45} S_{\QQ,\obss}(f,i) 
}
\end{eqnarray}%
}
where
\begin{eqnarray}
S_{\RR,\corr}(f,i) = 
S_{\RR,\obss}(f,i) -  
(\hat{a}_{0,\calcc}(i) -
\hat{\bar{a}}_{0,\calcc} ) 
S_{\QQ,\obss}(f,i).
\end{eqnarray}
In effect, given that our calibration
experiment measurement of $a_{0,\calcc}(i)$
has negligible systematic error,
the above correction 
returns produces a spectrum where
the values of $a_0$
should be nearly the same for all runs.
We stress that 
after selecting the 
model \textcolor{black}{based on cross-validation
analysis of corrected spectra (see Eq. 7)},
we estimate 
$a_0 - {\bar{a}}_{0,\calcc}$
from the uncorrected Eq. 6 spectrum.

In our cross-validation method,
we generate 20 000 random five-way splits of the data.
In each five-way split,
we assign the 
pair of spectra,
($S_{\RR,\corr}$
and $S_{\QQ,\obss})$,
from any particular run to one  of the five subsets 
by a
resampling
method.
Each of the 45 spectral pairs appears in one and only one
of the five subsets.
We resample spectra according to run to
retain possible correlation structure
within the spectrum for any particular run.
Each simulated five-way split is determined by
a random
permutation of the positive integers from 1 to 45.
The first nine permuted integers corresponds to the runs assigned to the
first subset. The second nine correspond to the runs assigned to the second subset, and so on.
From each random split,
four of the subsets are aggregated to form 
training data, and
the other subset forms the
validation data.
Within the training data, we pool all $S_{\RR,\corr}$ spectrum and
all 
$S_{\QQ,\obss}$ spectrum and form one ratio spectrum.
Similarly, for the validation data, we pool all $S_{\RR,\obss}$ spectrum and
$S_{\QQ,\obss}$ spectrum and form one ratio spectrum.
We fit each candidate polynomial ratio spectrum model to the training data, and predict
the observed ratio spectrum in the validation data
based on this fit.
We then compute
the (empirical)
mean squared deviation (MSD) between predicted and observed
ratios for the
validation data.
For any random five-way split, there are five
ways of defining the validation. Hence, we compute five MSD values for
each random split.
The cross-validation statistic for
each $d$,
CV($d$),
is the
average of these five MSD values.
For each random five-way split, we select the model
that yields the smallest value of CV($d$).
Based on 20 000 random splits of the 45 spectra,
we estimate
a probability mass function
$\hat{p}(d)$
where
the possible values of $d$ are:
2,4,6,8,10,12 or 14.

\subsection{Uncertainty quantification}
For any choice of $f_{\maxx}$, suppose that
$d$ is known exactly. For this ideal case,
based on a fit of the ratio spectrum model to the
Eq. 6 ratio spectrum,
we could construct a {coverage} interval for
$a_0$ with standard asymptotic methods or with
a parametric bootstrap method \cite{Ef2}.
{For our application,
we approximate the
parametric bootstrap
distribution} of our estimate of $a_0$ as
a Gaussian distribution with mean $\hat{a}_0$ and 
variance
$\hat{\sigma}^2_{\hat{a_0}(d),\rann}$,
where
$\hat{\sigma}_{\hat{a_0}(d),\rann}$
is predicted by asymptotic theory.  
To account for
the effect of
uncertainty
in $d$ on
our estimate of
$a_0$, we form a mixture of bootstrap distributions as follows
\begin{eqnarray}
f(x) =  \sum_{d}^{} \hat{p}(d) g( x, \hat{a}_0(d), \hat{\sigma}^2_{\hat{a_0}(d),\rann})
,
\end{eqnarray}
where
$g( x, \mu, \sigma^2)$ is the {probability density function} (pdf)
for a Gaussian random variable with expected value $\mu$ and
variance $\sigma^2$.
For any $f_{\maxx}$, we select the $d$
that yields the largest
value of
$\hat{p}(d)$.
Given that the probability density function (pdf) of a
random variable $X$ is $ f(x) = \sum_{i=1}^{n} w_i f_i(x)$,
and the mean and variance of a random variable $Z$
with pdf $f_i(z)$ are
$\mu_i$ and $\sigma^2_i$, the mean
and variance of $X$ are 
\begin{eqnarray}
E(X) = \mu =  \sum_{i=1}^{n} w_i \mu_i,
\end{eqnarray}
and
\begin{eqnarray}
VAR(X) =
\sum_{i=1}^{n} w_i \sigma^2_i 
+
\sum_{i=1}^{n} w_i  (\mu_i - \mu)^2.
\end{eqnarray}
Hence, the mean
and variance
of a random variable sampled from
the Eq. 9 pdf
are
$\hat{\bar{a}}_0 $
and
$\hat{\sigma}^2_{\tott}$
respectively, where
\begin{eqnarray}
\hat{ \bar{a}}_0  = 
\sum_{d} \hat{p}(d)  \hat{a}_0(d),
\end{eqnarray}
and
\begin{eqnarray}
\hat{\sigma}^2_{\tott}
=
\tilde{\sigma}^2_{\alpha}
+
\tilde{\sigma}^2_{\beta},
\end{eqnarray}
where
\begin{eqnarray}
\tilde{\sigma}^2_{\alpha} = \sum_{d} \hat{p}(d)
\hat{\sigma}^2_{\hat{a}_0(d),\rann},
\end{eqnarray}
and
\begin{eqnarray}
\tilde{\sigma}^2_{\beta} = \sum_{d} \hat{p}(d) ( \hat{a}_0(d) - \hat{\bar{a}}_0 ) ^ 2
.
\end{eqnarray}

For each $f_{\maxx}$ value,
we estimate the uncertainty of our estimate of
$a_0$ as 
$\hat{\sigma}_{\tott}$.
We select 
$f_{\maxx}$
by minimizing
$\hat{\sigma}_{\tott}$
on grid in frequency space.
For any fitting bandwidth,
$\tilde{\sigma}_{\beta}$
is the weighted-mean-square deviation
of the estimates of $a_0$ from the candidate models
about their weighted mean value where the weights
are the empirically determined selection fractions.
The term
$\tilde{\sigma}_{\alpha}$
is the weighted variance of
the parametric bootstrap sampling distributions for 
the candidate models where
the weights
are again the empirically determined selection fractions.
We stress that both
$\tilde{\sigma}_{\alpha}$
and
$\tilde{\sigma}_{\beta}$
are affected by
imperfect knowledge of the ratio spectrum model.

\section{Results}
\subsection{Analysis of Experimental data}

We fit candidate ratio spectrum models to the Eq. 6 observed ratio spectrum by the method of Least Squares (LS).
We determine model selection fractions (Table 1)
and $\hat{\sigma}_{\tott}$ (Table 2) for
$f_{\maxx}$ values on an evenly space grid
(with spacing of 25 kHz)
between
200 kHz and 1400 kHz.
In Figure 2,
we show how
selected $d$, $\hat{\sigma}_{\tott}$ and
$\hat{a}_0 - \hat{\bar{a}}_{0,\calcc}$ 
vary with $f_{\maxx}$.
Our method selects
$(f_{\maxx},d)$ = ( 1250 kHz, 8),
$\hat{\sigma}_{\tott,*}=$
3.25 $\times 10^{-6}$,
and
$\hat{a}_{0} - \hat{\bar{a}}_{0,\calcc} = $ 2.36
$\times 10^{-6}$.
In Figure 3, we show results for the values of
$f_{\maxx}$
(
that yield
the five lowest values of $\hat{\sigma}_{\tott}$.
For these five fitting bandwidths
(900 kHz, 1150 kHz,
1175 kHz, 1225 kHz and 1250 kHz)
$\hat{\sigma}_{\tott}$ values appear to follow no pattern
as a function of frequency, however
visual inspection suggests that the
\textcolor{black}{estimates
of
${a}_{0} - {\bar{a}}_{0,\calcc}$}
may follow a pattern.
It is not clear if the variations in Figure 2 and 3
are due to
random
or systematic measurement
effects.
To get some insight into this issue, we
study the performance of our method for
idealized simulated data that are free of systematic measurement error.

We simulate
three realizations of 
data based on the estimated values of 
$a_2,a_4,a_6 $ and $a_8$ in Table 3.
In our simulation, for each run, we set $a_0 - a_{0,\calcc}$ = 0,
${S_{\QQ,\obss}}(f)$ = 1
and 
${S_{\RR,\obss}}(f)$ equals the sum of the predicted
ratio spectrum and Gaussian white noise.
For each run,
the variance of the noise
is 
determined from 
fitting
the ratio spectrum model to the experimental ratio spectrum for that run.
For simulated data, the
$d$, $\hat{\sigma}_{\tott}$
and $\hat{a}_0 - a_{0,\calcc}$ spectra 
exhibit fluctuations 
similar to those in the experimental data (Figures 4,5,6).
Since
variability
in simulated data
(Figures 4,5,6)
is due solely to random effects,
we can not rule out the possibility that
random effects may have produced
the
fluctuations in the
experimental spectra (Figures 2 and 3).
For the third realization of
simulated data (Figure 6), the estimated values of $a_0 - a_{0,\calcc}$ 
corresponding to the $f_{\maxx}$ values that yield
the five lowest values of $\hat{\sigma}_{\tott}$ 
form two clusters in $f_{\maxx}$ space which are separated by
approximately 300 kHz (Figure 7).
This pattern is similar
to that 
observed for the experimental data (Figures 3).
As a caveat,
for the experimental data, we can not rule
out the possibility that systematic measurement error could
cause or enhance observed
fluctuations.


Our method correctly selects the $d$=8
model for each of three independent realizations of simulation data (Table 4).
In a second study, we simulate
three realizations of
data according to a $d$=6 polynomial model
based on  the fit to experimental 
data for $f_{\maxx} = $ 900 kHz.
Our methods correctly selects the $d$=6
for each of the three realizations.

In \cite{Qu01},
our method selected
$f_{\maxx}=$ 575 kHz and $d=4$
when $f_{\maxx}$ was constrained to be less than
600 kHz.
For these selected values,
our current analysis yields
$\hat{a}_0 - \hat{\bar{a}}_{0,\calcc} = $ 1.81
$\times 10^{-6}$
and   
$\hat{\sigma}_{\tott}=$ 3.58
$\times 10^{-6}$.
In this study,
when $f_{\maxx}$ is allowed to be as large as 1400 kHz,
our method selects
$f_{\maxx}$ = 1250 Hz and $d=8$,
and
$\hat{a}_0 - \bar{\hat{a}}_{0,\calcc} = $ 2.36
$\times 10^{-6}$
and
$\hat{\sigma}_{\tott,*}=$ 3.25
$\times 10^{-6}$.
The difference between
the two results,
0.55 
$\times 10^{-6}$,
is small compared to the
uncertainty of either result.

We expect imperfections in our fitting
bandwidth selection method for various reasons.
First, 
we conduct a grid search with a resolution of 25 kHz
rather than a search over a continuum of frequencies.
Second, there are 
surely
fluctuations in 
$\hat{\sigma}_{\tott}$
due to random effects that vary with fitting bandwidth.
Third,
different values of
$f_{\maxx}$ can yield very similar values of
$\hat{\sigma}_{\tott}$ but somewhat different values of
$\hat{a}_0 - \hat{\bar{a}}_{0,\calcc}$.
Therefore, it is reasonable to 
determine an additional component of uncertainty, $\hat{\sigma}_{f_{\maxx}}$,
that accounts for uncertainty due to imperfect
\textcolor{black}{performance of our method to
select $f_{\maxx}$.}
Here we equate
$\hat{\sigma}_{f_{\maxx}}$
to the estimated standard deviation of
estimates of
$a_0  -{\bar{a}}_{0,\calcc}$ 
corresponding to the
five $f_{\maxx}$ values that yielded
the lowest $\hat{\sigma}_{\tott}$ values (Figures 3 and 7).
For the three realizations of simulated spectra, the corresponding
$\hat{\sigma}_{f_{\maxx}}$
values {are}
0.39 $\times 10^{-6}$,
0.45 $\times 10^{-6}$,
and
1.83 
 $\times 10^{-6}$.
For the experimental data
$\hat{\sigma}_{f_{\maxx}}= $ 0.56
$\times 10^{-6}$.
For both simulated and experimental data, our update for
the total uncertainty of estimated  $a_0$
is
$\hat{\sigma}_{\tott,\finall}$ where  
\begin{eqnarray}
\hat{\sigma}_{\tott,\finall}  = 
\sqrt{ 
\hat{\sigma^2}_{\tott,*}  + 
\hat{\sigma^2}_{f_{\maxx}} }.
\end{eqnarray}
For the simulated data,
$\hat{\sigma}_{\tott,\finall}$
is
2.69 $\times 10^{-6}$,
3.08 $\times 10^{-6}$,
and
3.42 
$\times 10^{-6}$.
For the experimental data,
$\hat{\sigma}_{\tott,\finall}= $   
3.29
$\times 10^{-6}$.
As a caveat, the choice to quantify 
$\hat{\sigma}_{f_{\maxx}}$ as the standard deviation
of the estimates corresponding to $f_{\maxx}$ values that
yield the five lowest values is
based on scientific judgement.
For instance, 
if we determine
$\hat{\sigma^2}_{f_{\maxx}} $ based on the
the $f_{\maxx}$ values that yield the ten lowest
rather the five lowest
values $\hat{\sigma}_{\tott}$,
for the three simulation cases and the observed data
we get
$\hat{\sigma}_{f_{\maxx}}  =$ 
0.55 $\times 10^{-6}$,
1.08 $\times 10^{-6}$,
1.42 $\times 10^{-6}$,
and
0.73 $\times 10^{-6}$ respectively.

\subsection{Stability analysis}
From the corrected Eq.7 spectra,
we estimate
$a_0 - \bar{a}_{0,\calcc}$
for each of the 45 runs
by
fitting our selected ratio spectrum model
($d=$ 8 and $f_{\maxx}=$ 1250 kHz)
to the data from each
run
by the method of LS
(Figure 8).
Ideally, on average, these
estimates should not \textcolor{black}{vary} from run-to-run assuming that
our Eq. 8 correction model based on calibration 
data is valid.
From these estimates, we determine the
slope and intercept
parameters for a linear trend
model by the method of Weighted Least Squares (WLS)
where we minimize
\begin{eqnarray}
\chi^2_{\obss} 
=
\sum_{i=1}^{45} w_i ( y_i - \hat{y}_i )^2.
\end{eqnarray}
Above, for the $i$th run,
$y_i$ and $\hat{y}_i$
are the estimated and predicted (according to the trend model) 
values of
$a_{0} - {\bar{a}}_{0,\calcc}$,
and
$w_i$ is the 
inverse of the squared
asymptotic uncertainty of
associated with our estimate
determined by the LS fit to data from the  $i$th run.
We determine the uncertainty of the trend model
parameters with a nonparametric bootstrap method (see Appendix 1)
following \cite{DH}
(Table 5).
We repeat the bootstrap procedure but set 
$\hat{y_i}$ to a constant.
This analysis yields an estimate of the null distribution 
of the slope estimate
corresponding to the hypothesis that there is no trend.
The fraction of bootstrap slope estimates with magnitude greater than
or equal to the magnitude of the estimated slope determined
from the observed data is the bootstrap $p$-value
\cite{Ef2}
corresponding to a two-tailed test of the null hypothesis.
For the $f_{\maxx}$ values that yield
the five lowest values of $\hat{\sigma}_{tot}$,
our bootstrap analysis provides strong evidence that
$a_{0} - {\bar{a}}_{0,\calcc}$ varies with time.
At the value of $f_{\maxx} = $ 575 kHz,
there is a moderate amount of evidence for a trend.

For each $f_{\maxx}$ choice,
we test
the hypothesis that the linear trend is consistent
with observations based on the value of
$\chi^2_{\obss}$.
If the observed data are consistent with the
trend model,
a resulting $p$-value
from this test
is realization of a random variable
with a distribution that is approximately uniform between 0 and 1.
Hence, the large $p$-values reported column 7 of Table 5 suggest that
the asymptotic uncertainties
determined by the LS method
for each run may be inflated.
As a check, we estimate the slope uncertainty with a 
parametric bootstrap method where we simulate
bootstrap replications of the observed data by adding
Gaussian noise to the estimated trend with standard deviations equal to the
asymptotic uncertainties determined by the 
LS method.
In contrast to the method from
\cite{DH},
the
parametric bootstrap method yields larger slope uncertainties.
For instance, for the 1250 kHz and 575 kHz cases, the parametric
bootstrap slope uncertainty estimates are larger than the
corresponding Table 5 estimates by 30 $\%$ and 
23 $\%$ respectively.
\textcolor{black}{That is, the parametric bootstrap method estimates
are inflated with respect to the estimates determined with the
method from $\cite{DH}$.}

\textcolor{black}{This inflation} 
could result due to
heteroscedasticity
(frequency dependent additive measurement error variances).
This follows from the well-known fact that when
models are fit to heteroscedastic data, the variance of
parameter estimates determined by the LS method
are larger than the variance of parameter estimates determined
by the ideal WLS method.
Based on fits of selected models to data pooled from all runs,
we test the hypothesis that the variance of the additive noise
in the ratio spectrum is independent of frequency.
Based on
the
Breush-Pagan method \cite{BP}, the $p$-values
corresponding to the test of this hypothesis are
0.723, 0.064, 0.006, 0.001, 0.001 and 0.001 for 
fitting bandwidths of 575 kHz, 
900 kHz, 
1150 kHz, 
1175 kHz, 
1225 kHz, 
and 1250 kHz respectively.
Hence, the evidence for heteroscedasticity is
very
strong for the larger fitting bandwidths.

For a fitting bandwidth of 1250 kHz,
the variation of the estimated trend
over the duration of the experiment
is
$18.2 \times 10^{-6}$.
In contrast, 
the uncertainty of
our estimate
of
$a_{0} - \bar{a}_{0,\calcc}$ 
determined under the assumption that there is no trend
is only
3.29
$\times 10^{-6}$ (Table 4).
The  evidence for a linear trend at particular fitting bandwidths above 900 kHz
is strong (p-values are 0.012 or less) (column 5 in Table 5).
In contrast, the evidence (from hypothesis testing) for a linear trend at 575 kHz
(the selected fitting bandwidth in \cite{Qu01})
is not as strong since
the p-value is 0.049.
However, the larger p-value at 575 kHz may be due to more random variability
in the estimated offset parameters rather than lack of a deterministic trend 
in the offset parameters.
This hypothesis is supported by the
observation that the uncertainty of the estimated slope parameter
due to random effects generally increases as the fitting bandwidth
is reduced, and
that fact that
all of the slope parameter
estimates for cases
shown in Table 5 are negative
and vary from
-0.440 $\times 10^{-6}$ / day
to
-0.164 $\times 10^{-6}$ / day.
Together,
these observations strongly suggest a
deterministic trend in measured offset parameters at a fitting bandwidth of 575 kHz.
Currently, experimental efforts are underway to understand the physical source
of this (possible) temporal trend.

If there is
a linear temporal trend in the offset parameters at a fitting bandwidth of 575 kHz,
with slope similar to what we estimate from data,
the reported uncertainty for the Boltzmann constant reported in
\cite{Qu01} is optimistic because the trend was not accounted for
in the uncertainty budget.
To account for the effect of a linear trend on results, one must
estimate the associated systematic
error due to the trend at some particular time. 
Unfortunately,
we have no empirical method to
do this.
As a further complication, 
if there is a trend, it may not be exactly linear.

\section{Summary}
For electronic measurements of the Boltzmann constant
by JNT, we
presented a data-driven method
to select the complexity of
a ratio spectrum  model
with 
a cross-validation
method.
For each candidate fitting bandwidth,
we quantified
the
uncertainty of
the offset parameter in the spectral ratio model
in a way that
accounts for random effects as well as systematic
effects associated with imperfect knowledge of model 
complexity.
We selected the fitting bandwidth that minimizes
this uncertainty.
We also quantified an additional component of uncertainty
that accounts for imperfect 
knowledge of the selected fitting bandwidth.
With our method, we re-analyzed data from a recent experiment by considering a broader
range of fitting bandwidths and found evidence for a temporal linear trend
in offset paramaters.
For idealized simulated data
free of systematic error with additive noise
similar  
to  experimental data,
our method correctly selected the true complexity
of the ratio spectrum model for all
cases considered.
In the future, we plan to study
how well our methods work for other experimental and simulated JNT data sets
with variable signal-to-noise ratios and investigate how
robust our method is to systematic measurement errors.
We expect our method to find broad
metrological applications including
selection of optimal virial equation models
in gas thermometry experiments,
and selection of line-width models in Doppler broadening thermometry experiments.

{\bf Acknowledgements.}
Contributions of staff from NIST, an agency of the US government,
are not subject to copyright in the US.
JNT research at NIM is supported by grants NSFC (61372041 and 61001034).
Jifeng Qu acknowledges Samuel P. Benz, Alessio Pollarolo, Horst Rogalla,
Weston L. Tew of the NIST and Rod White of MSL, New Zealand for
their help with
the JNT measurements analyzed in this work. 
We also thank Tim Hesterberg of Google for 
helpful discussions.

\section{Appendix 1: nonparametric bootstrap resampling method}
We denote the estimate of
$a_0 - \bar{a}_{0,\calcc}$
from the $i$th run as
$y_i$,
and the data acquisition time for the $i$th run as $t_i$.
Our linear trend model
is
\begin{eqnarray}
y = X \beta + \epsilon,
\end{eqnarray}
where the observed data is $y=(y_1,y_2 \cdots y_{45})^T$,
$\epsilon$
is an additive noise
term,
and
the matrix
$X$ is
 \[ \left( \begin{array}{cc}
X_{1~1} & X_{1~2}\\
X_{2~1} & X_{2~2}\\
. & . \\
. & . \\
. & . \\
X_{45~1} & X_{45~2} \end{array} \right) 
=
  \left( \begin{array}{cc}
1 & 0   \\
1 & t_2 - t_1  \\
. & .  \\
. & .  \\
. & .  \\
1 & t_{45} - t_1  \end{array} \right),\]
and ${\beta} = {( \beta_0 , \beta_1 )}^{T}$ where
the $\beta_0$ is the intercept parameter and $\beta_1$ is the slope parameter.
Above, we model
the $i$th component of
$\epsilon$ as a realization of a random variable with 
expected value 0 and variance $ \kappa V_i$
where $V_i$ is known but $\kappa$ is unknown.
Following \cite{DH},
the predicted value 
in a linear regression model is $\hat{y}= X \hat{\beta}$
where the estimated model parameters  $\hat{\beta}$
are
\begin{eqnarray}
\hat{\beta} = ( X^T W X )^{-1} X^T W y,
\end{eqnarray}
where $W$ is a diagonal weight matrix.
Here, we set the $j$th component of $W$
to $w_j =  \frac{1}{V_j}$ where
$V_j$ is the 
estimated asymptotic variance of $y_i$ determined by the method of LS.
Following \cite{DH}, we form a modified residual
\begin{eqnarray}
r_i = \frac{  y_i - \hat{y}_i } {  \sqrt{V_i( 1 - h_i) } },
\end{eqnarray}
where $h_i$ is the i$th$ diagonal element of the
hat matrix $H = X( X^T W X)^{-1}X^T W$. This transformation ensures that
the modified residuals are realizations of random variables with
nearly the same variance.  
The $i$th component of
a bootstrap replication of the 
observed data is
\begin{eqnarray}
y^{*}_i = \hat{\beta}_0 + \hat{\beta}_1 (t_i - t_1) + \sqrt{V_i}  {e^{*}_i},
\end{eqnarray}
where $i =1,2, \cdots 45$ and
$(e^{*}_1, e^{*}_2, \cdots e^{*}_{45}) $
is sampled
with replacement
from  $(r_1 - \bar{r}, r_2 - \bar{r}, \cdots r_{45} - \bar{r})$ 
where $\bar{r}$ is the mean modified residual.
From each of 50 000 bootstrap replications of the observed data,
we determine a slope and intercept parameter.
The bootstrap estimate of the uncertainty of the
slope and intercept parameters are the estimated standard
deviations of the slope and intercept parameters determined
from  these estimates.

\section{Appendix 2: Summary of model selection and uncertainty quantification method}

\noindent
1. Based on a calibration data estimate of $a_{0,\calcc}$ for each run, correct
$S_{\RR,\obss}$ spectra per Eq. 8.
\\

\noindent
2. Set fitting bandwidth to $f_{\maxx}$ = 200 kHz.
\\

\noindent
3. Set resampling counter to $n_{\resampp}=$ 1.
\\

\noindent
4. Randomly split
$S_{\RR,cor}$ and $S_{\QQ,\obss}$ spectral pairs  
from each of 45 runs
into five subsets.
Select $d$ by five-fold cross-validation
and
update appropriate model selection fraction estimate.
\\

\noindent
5. Increase $n_{\resampp}$ by 1.
\\

\noindent
6.  If $n_{\resampp} < $ 20 000, go to 4.
\\

\noindent
7.  If $n_{\resampp} = $ 20 000,
select the polynomial model with largest associated selection fraction.
Calculate estimate of residual offset and its uncertainty
$\hat{\sigma}_{\tott,\finall}$ (Eq. 16) from pooled uncorrected spectrum (see Eq. 6).
\\

\noindent
8. Increase $f_{\maxx}$ by 25 kHz
and go to 3 if $f_{\maxx} < $ 1400 kHz.
\\

\noindent
9. Select
$f_{\maxx}$
by minimizing
$\hat{\sigma}_{\tott}$. Denote the minimum value as
$\hat{\sigma}_{\tott,*}$.
\\

\noindent
10. Assign component of uncertainty associated with imperfect
performance of method to select $f_{\maxx}$ as
$\hat{\sigma}_{f_{\maxx}}$ to estimated
standard deviation of
offset estimates
corresponding to $f_{\maxx}$ values that yield
the five lowest values of 
$\hat{\sigma}_{\tott}$. 
Our final estimate of the uncertainty of the offset is
$\hat{\sigma}_{\tott,\finall} = \sqrt{  
\hat{\sigma^2}_{\tott,*} +
\hat{\sigma^2}_{f_{\maxx}}}$.
\\

For completeness, we note that
simulations and analyses were
done with
software scripts
developed in the public domain R \cite{RR} language and environment.
Please contact Kevin Coakley regarding any software-related questions.

\newpage{}

\begin{table}
\centering
\begin{tabular}{c|ccccccc}
\hline
$f_{\maxx}$ & selection fractions: $p(d)$ &&&  \\
(kHz) & $d$=2 & $d$=4 & $d$=6 & $d$=8 & $d$=10 & $d$=12 & $d$=14  \\
\hline\hline
\\
  200& 0.7789& 0.1269& 0.0698& 0.0046& 0.0159& 0.0039& 0.0000
\\
  300& 0.3271& 0.2510& 0.0000& 0.4160& 0.0047& 0.0012& 0.0000
\\
  400& 0.0304& 0.2234& 0.6427& 0.0017& 0.0240& 0.0716& 0.0061
\\
  500& 0.0249& 0.6043& 0.0010& 0.2648& 0.0701& 0.0001& 0.0349
\\
  525& 0.1347& 0.0960& 0.0150& 0.6758& 0.0747& 0.0037& 0.0000
\\
  550& 0.0290& 0.7016& 0.0000& 0.0000& 0.2668& 0.0027& 0.0000
\\
  575& 0.0000& 0.9446& 0.0000& 0.0000& 0.0012& 0.0513& 0.0029
\\
  600& 0.0000& 0.9264& 0.0003& 0.0000& 0.0002& 0.0730& 0.0000
\\
  625& 0.0027& 0.2192& 0.1273& 0.0746& 0.0002& 0.5289& 0.0472
\\
  650& 0.1072& 0.0022& 0.2342& 0.1557& 0.0006& 0.0000& 0.5001
\\
  675& 0.0320& 0.0287& 0.6831& 0.0077& 0.0201& 0.0000& 0.2285
\\
  700& 0.2659& 0.0106& 0.5514& 0.0002& 0.1718& 0.0000& 0.0002
\\
  725& 0.0326& 0.0181& 0.8166& 0.0159& 0.1067& 0.0101& 0.0000
\\
  750& 0.0700& 0.0103& 0.5985& 0.1475& 0.0242& 0.1495& 0.0001
\\
  775& 0.0000& 0.0863& 0.2889& 0.3968& 0.1279& 0.0980& 0.0021
\\
  800& 0.0000& 0.0216& 0.4607& 0.4439& 0.0017& 0.0600& 0.0120
\\
  825& 0.0009& 0.0667& 0.8472& 0.0320& 0.0004& 0.0168& 0.0362
\\
  850& 0.0000& 0.0000& 0.9294& 0.0071& 0.0034& 0.0059& 0.0541
\\
  875& 0.0000& 0.0000& 0.8301& 0.0344& 0.0001& 0.0000& 0.1354
\\
  900& 0.0278& 0.0012& 0.9357& 0.0314& 0.0030& 0.0000& 0.0009
\\
  925& 0.0178& 0.0002& 0.6384& 0.3411& 0.0003& 0.0000& 0.0022
\\
  950& 0.1015& 0.0000& 0.0135& 0.8824& 0.0022& 0.0006& 0.0000
\\
  975& 0.0421& 0.0212& 0.0782& 0.8560& 0.0026& 0.0000& 0.0000
\\
 1000& 0.0000& 0.1072& 0.1866& 0.6441& 0.0622& 0.0000& 0.0000
\\
 1025& 0.0000& 0.1704& 0.0000& 0.8158& 0.0074& 0.0029& 0.0035
\\
 1050& 0.0010& 0.0000& 0.0000& 0.9751& 0.0232& 0.0007& 0.0000
\\
 1075& 0.0117& 0.0001& 0.0000& 0.9646& 0.0235& 0.0000& 0.0000
\\
 1100& 0.0000& 0.0021& 0.0000& 0.9976& 0.0003& 0.0000& 0.0000
\\
 1125& 0.0000& 0.0000& 0.0000& 0.9519& 0.0481& 0.0000& 0.0000
\\
 1150& 0.0000& 0.0000& 0.0000& 0.9962& 0.0012& 0.0016& 0.0010
\\
 1175& 0.0000& 0.0000& 0.0000& 0.9996& 0.0000& 0.0003& 0.0000
\\
 1200& 0.0000& 0.0002& 0.0000& 0.7125& 0.1578& 0.1269& 0.0027
\\
 1225& 0.0000& 0.0000& 0.0000& 0.9998& 0.0002& 0.0000& 0.0000
\\
 1250& 0.0000& 0.0000& 0.0000& 0.9427& 0.0571& 0.0003& 0.0000
\\
 1275& 0.0000& 0.0000& 0.0000& 0.4784& 0.5181& 0.0034& 0.0000
\\
 1300& 0.0000& 0.0000& 0.0000& 0.1782& 0.5986& 0.2231& 0.0000
\\
 1325& 0.0000& 0.0000& 0.0000& 0.0010& 0.1050& 0.8885& 0.0055
\\
 1350& 0.1879& 0.2538& 0.0000& 0.0000& 0.0000& 0.2315& 0.3268
\\
 1375& 0.0362& 0.0322& 0.0000& 0.0000& 0.1788& 0.7038& 0.0491
\\
 1400& 0.2812& 0.0593& 0.0000& 0.0000& 0.4527& 0.0060& 0.2007
\\
\hline
\end{tabular}
\caption{
Moodel selection
fractions for polynomial models for experimental data.
}
\label{table:examples}
\end{table}
\newpage{}

\begin{table}
\centering
\begin{tabular}{cccccccc}
\hline
$f_{\maxx}$ & $d$ & $\hat{a}_0 - \hat{\bar{a}}_{0,\calcc}$  & $\hat{\sigma}_{\hat{a}_0(d),\rann} $  &$\hat{\bar{a}}_0 -\hat{\bar{a}}_{0,\calcc} $ &$\tilde{\sigma}_{\alpha}$ & $\tilde{\sigma}_{\beta} $ & $\hat{\sigma}_{\tott}$ \\
\\
(kHz)& & $\times 10^6 $  & $\times 10^6 $  & $\times 10^6$ &$\times 10^6$ & $\times 10^6$ & $\times 10^6$ \\
\hline\hline
  200&    2&  -0.57&   4.15&  -1.97&   4.57&  3.173&  5.564
\\
  300&    8&  -7.58&   5.62&  -2.71&   4.67&  4.348&  6.380
\\
  400&    6&  -2.09&   4.40&  -1.35&   4.41&  3.041&  5.357
\\
  500&    4&   1.88&   3.47&   0.39&   4.00&  2.133&  4.535
\\
  525&    8&  -0.42&   4.38&  -0.75&   4.13&  1.255&  4.319
\\
  550&    4&   1.90&   3.26&   0.51&   3.69&  2.179&  4.290
\\
  575&    4&   1.81&   3.18&   1.48&   3.31&  1.360&  3.579
\\
  600&    4&   2.21&   3.14&   1.79&   3.30&  1.473&  3.618
\\
  625&   12&  -2.65&   4.83&  -1.04&   4.31&  2.295&  4.886
\\
  650&   14&  -5.09&   5.05&  -2.35&   4.34&  4.384&  6.167
\\
  675&    6&   3.55&   3.49&   1.36&   3.88&  3.455&  5.197
\\
  700&    6&   2.88&   3.40&  -0.95&   3.32&  5.318&  6.271
\\
  725&    6&   2.63&   3.37&   1.92&   3.46&  2.345&  4.177
\\
  750&    6&   1.99&   3.39&   0.97&   3.61&  3.270&  4.874
\\
  775&    8&   3.88&   3.72&   1.78&   3.68&  2.422&  4.404
\\
  800&    6&   1.15&   3.29&   1.96&   3.57&  1.699&  3.953
\\
  825&    6&   1.64&   3.31&   0.92&   3.39&  2.454&  4.182
\\
  850&    6&   1.61&   3.25&   1.50&   3.36&  0.545&  3.403
\\
  875&    6&   1.29&   3.21&   1.07&   3.45&  0.735&  3.530
\\
  900&    6&   1.44&   3.17&   1.36&   3.17&  0.769&  3.266
\\
  925&    6&   1.03&   3.13&   1.49&   3.27&  0.656&  3.334
\\
  950&    8&   2.85&   3.51&   3.12&   3.44&  0.994&  3.577
\\
  975&    8&   2.35&   3.44&   2.04&   3.39&  4.020&  5.261
\\
 1000&    8&   1.97&   3.41&  -1.29&   3.33&  8.390&  9.028
\\
 1025&    8&   2.73&   3.44&  -2.49&   3.41& 11.510& 12.010
\\
 1050&    8&   2.90&   3.40&   2.91&   3.41&  0.968&  3.549
\\
 1075&    8&   2.94&   3.37&   3.38&   3.40&  4.185&  5.394
\\
 1100&    8&   2.78&   3.37&   2.69&   3.37&  1.842&  3.837
\\
 1125&    8&   3.13&   3.34&   3.09&   3.36&  0.343&  3.372
\\
 1150&    8&   2.69&   3.30&   2.69&   3.30&  0.329&  3.315
\\
 1175&    8&   2.82&   3.30&   2.82&   3.30&  0.012&  3.301
\\
 1200&    8&   2.18&   3.28&   2.34&   3.42&  0.780&  3.511
\\
 1225&    8&   2.62&   3.25&   2.62&   3.25&  0.002&  3.253
\\
 1250&    8&   2.36&   3.22&   2.40&   3.24&  0.145&  3.246
\\
 1275&   10&   3.13&   3.53&   2.56&   3.38&  0.592&  3.431
\\
 1300&   10&   3.52&   3.48&   2.88&   3.49&  0.863&  3.597
\\
 1325&   12&   2.23&   3.76&   2.42&   3.73&  0.553&  3.774
\\
 1350&   14&   3.38&   4.05&  25.47&   6.62& 90.27& 90.52
\\
 1375&   12&   2.82&   3.81&   9.22&   4.54& 43.23& 43.47
\\
 1400&   10&   4.32&   3.57&  68.13&   8.41&112.2&112.5
\\
\hline
\end{tabular}
\caption{
Results for experimental data.}
\label{table:thresholds}
\end{table}
\newpage{}

\begin{table}
\centering
\begin{tabular}{cc}
\hline
parameter & estimate
\\
\\
$a_0 - {\bar{a}}_{0,\calcc}$ &   2.36 (3.22) $\times 10^{-6}$
\\
$a_2$ & -4.33 (0.41) $\times 10^{-4}$
\\
$a_4$ & 1.66 (0.13) $\times 10^{-3}$
\\
$a_6$ & -2.25 (0.13) $\times 10^{-3}$
\\
$a_8$ &  6.26 (0.46) $\times 10^{-4}$
\\
\hline
\end{tabular}
\caption{
Estimated model parameters 
for experimental data for
$f_{\maxx} = $ 1250 kHz.
}
\label{table:sim}
\end{table}

\begin{table}
\centering
\begin{tabular}{cccccccccccc}
\hline
& $f_{\maxx}$ & $d$ & $\hat{a}_0 - \hat{\bar{a}}_{0,\calcc}$  & $\hat{\sigma}_{\hat{a}_0(d),\rann} $  &$\hat{\bar{a}}_0 -\hat{\bar{a}}_{0,\calcc} $ &$\tilde{\sigma}_{\alpha}$ & $\tilde{\sigma}_{\beta} $ & 
$\hat{\sigma}_{\tott,*}$ 
&
$\hat{\sigma}_{f_{\maxx}}$ 
&
$\hat{\sigma}_{\tott,\finall}$ 
\\
& (kHz) &  & $\times 10^6 $  & $\times 10^6 $  & $\times 10^6$ &$\times 10^6$ & $\times 10^6$ & 
$\times 10^6$ & 
$\times 10^6$ & 
$\times 10^6$
\\
\\
\hline\hline
\\
\\
experimental& 1250 & 8 &  2.36 &  3.22 &  2.40  & 3.24 & 0.14 &  3.25 & 0.56
& 3.29
\\
\\
\\
realization 1 & 1400 & 8 & 4.25 & 2.64 & 4.20 &  2.66 & 0.18 & 2.67 & 0.39
& 2.69
\\
realization 2 & 1150 & 8 & -0.43 & 2.98 &  -0.39 & 3.00 &  0.59 & 3.05 & 0.45
& 3.08
\\
realization 3 & 1325 & 8 & -2.59 & 2.76 &  -2.88 & 2.82  &  0.60 &  2.89 & 1.83 
& 3.42
\\
\\
\hline
\end{tabular}
\caption{
Comparison of selected $f_{\maxx}$
and associated model parameters for
experimental and 
simulated data sets.
In the simulation, the true values of 
${a}_0 - a_{0,\calcc}$ 
and $d$ are 0 and 8 respectively.
The component of uncertainty $\hat{\sigma}_{f_{\maxx}}$
accounts for imperfect
performance of our
$f_{\maxx}$
selection
method.
}
\label{table:sim}
\end{table}
\newpage{}

\begin{table}
\centering
\begin{tabular}{cccccccc}
\hline
$f_{\maxx}$ (kHz) & $d$ & intercept & 
slope & $p$-value & $\chi^2_{\obss}$ & $p$-value & ${a}_0 - {\bar{a}}_{0,\calcc}$ 
\\
& & $\times 10^{6}$ &
$\times 10^{6}$ day  
 & (trend test) & & (model consistency test)& 
$\times 10^{6}$
\\
\hline\hline
\\
200 & 2 & 11.98(6.50) & -0.243(0.125) & 0.050 &  33.0 & 0.864 &  -0.57(5.56)
\\
300 & 8 & 11.74(10.54) & -0.440(0.199) & 0.027 & 47.0 & 0.311 &  -7.58(6.38)
\\
400 & 6 & 10.31(6.64) & -0.247(0.126) & 0.062 &  35.8 & 0.775 &  -2.09(5.36)
\\
500 & 4 & 10.69(5.12) & -0.175(0.097) & 0.072 &  32.6 & 0.877 &  1.88(4.54)
\\
575 & 4 & 10.13 (4.42) & -0.164 (0.084) & 0.049 & 28.2 & 0.960 
& 1.81(3.58)
\\
700 & 6 & 13.56(4.95) & -0.213(0.093) & 0.021 &  31.0 & 0.914 &  2.88(6.27)
\\
800 & 6 & 10.49(3.94) & -0.175(0.074) & 0.018 &  22.5 & 0.996 &  1.15(3.95)
\\
\\
\\
900 & 6 & 9.15 (3.49) & -0.164 (0.066) & 0.012 & 19.9 & 0.999
& 1.44(3.27)
\\
1150 & 8 & 11.39 (3.79) & -0.179 (0.071) & 0.011 & 23.2 & 0.994
& 2.69(3.32)
\\
1175 & 8 & 11.57 (3.81) & -0.185 (0.072) & 0.010 & 24.0 & 0.991
& 2.82(3.30)
\\
1225 & 8 & 12.13 (3.79) & -0.204 (0.071) & 0.004 & 25.0 & 0.987 &
2.62(3.25)
\\
1250 & 8 & 12.02 (3.78) & -0.202 (0.071) & 0.004 & 25.2 & 0.986
& 2.36(3.25)
\\ 
\hline
\end{tabular}
\caption{
Estimated trend model parameters,
associated uncertainties and $p$-values
for testing the no-trend hypothesis (column 5).
Based on 
$\chi^2_{\obss}$
and 
the number of degrees-of-freedom
(43),
we determine a $p$-value (column 7) for testing
the hypothesis that the linear trend
model is consistent with observations.
The value
$f_{\maxx} = $ 575 kHz corresponds to 
selected fitting bandwidth in \cite{Qu01}.
The five values (ranging from
900 kHz to 1250 kHz)
yield the five lowest values of
$\hat{\sigma}_{\tott}$
when $f_{\maxx}$ is varied from
200 kHz to 1400 kHz on a grid.
For comparison, we list results
at
fitting bandwidths of 200 kHz, 300 kHz, 400 kHz,
500 kHz, 700 kHz and 800 kHz.
In the last column, we list estimates of
$a_0 - \bar{a}_{0,\calcc}$ from the pooled data 
and their associated $\hat{\sigma}_{\tott}$ values in parentheses.}
(Table 2). 
\label{table:sim}
\end{table}
\newpage{}

\begin{figure}[!t]
\centering
\includegraphics[height=7.5in]{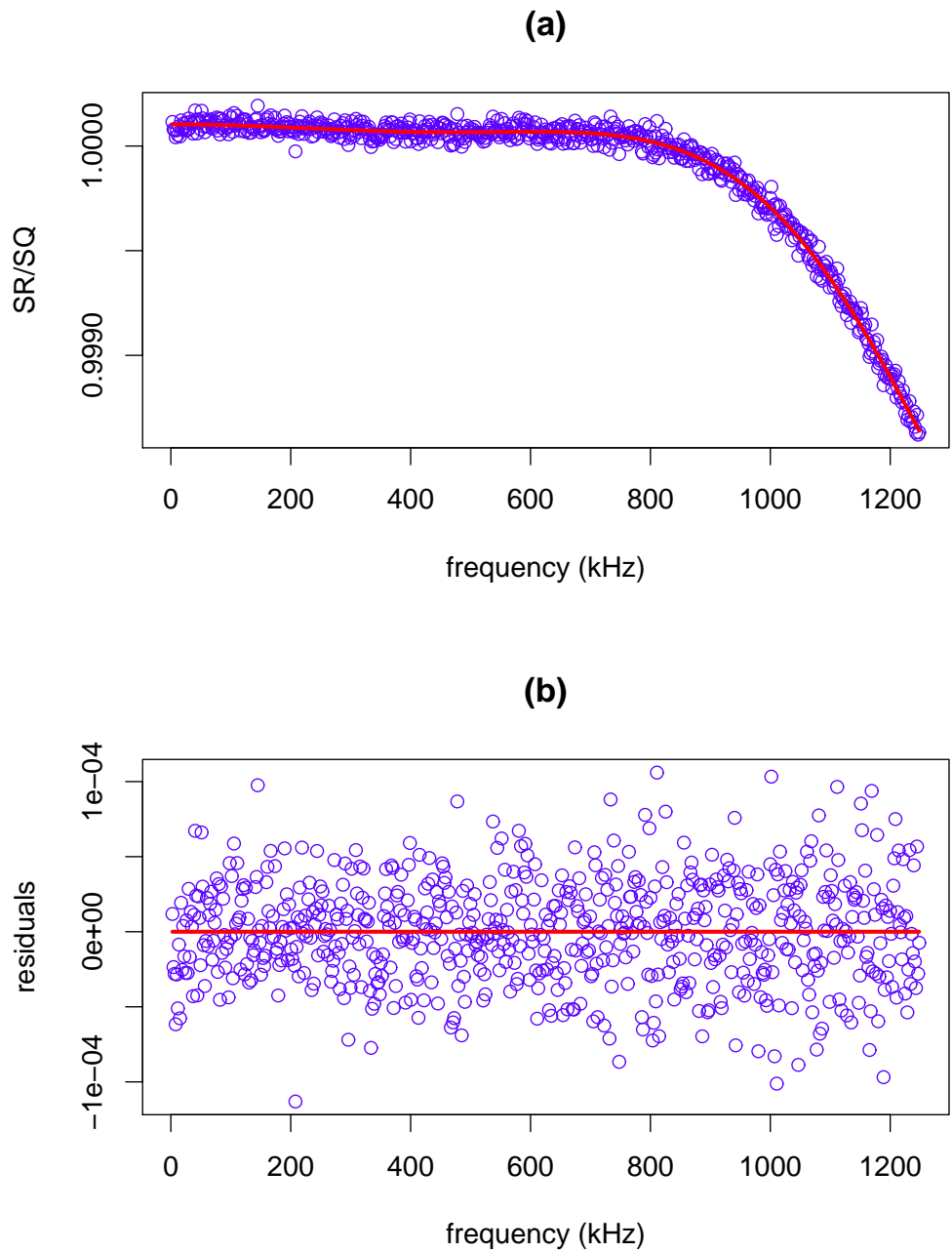}
\caption{Experimental data.
(a) Observed (Eq. 6) and predicted ratio spectra
based on selected model ($d$=8) and fitting bandwidth (1250 kHz)
(see Table 3).
(b) Residuals (observed - predicted).
}
\label{fig1}
\end{figure}
\clearpage{}

\begin{figure}[!t]
\centering
\includegraphics[height=7.5in]{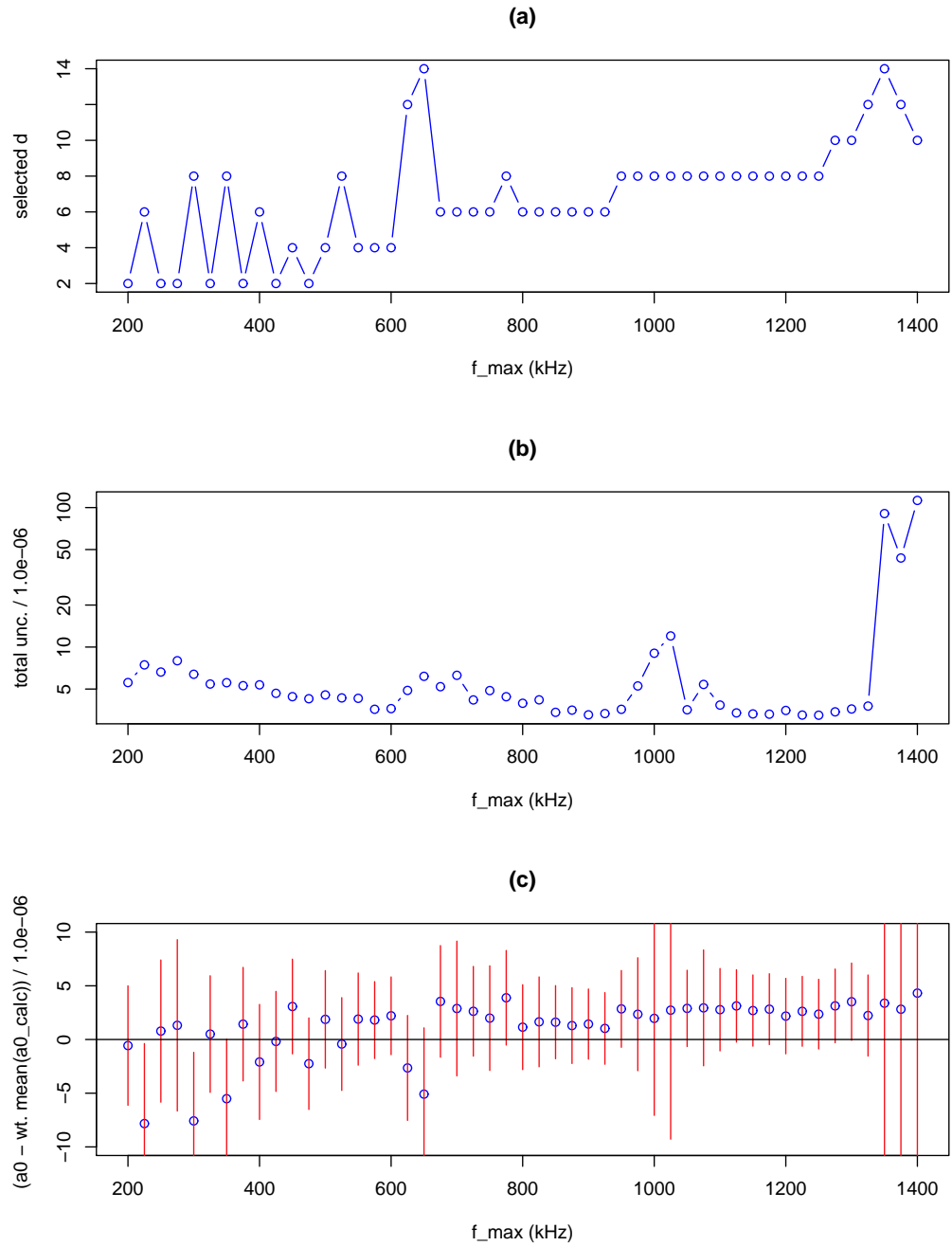}
\caption{Experimental data. (a) Estimated polynomial complexity parameter $d$.
(b) Estimated total uncertainty $ \hat{\sigma}_{\tott}$ (Eq. 13).
(c) Estimated $a_0 - \bar{a}_{0,\calcc}$ and approximate 68 $\%$ coverage interval.
}
\label{fig2}
\end{figure}

\begin{figure}[!t]
\centering
\includegraphics[height=7.5in]{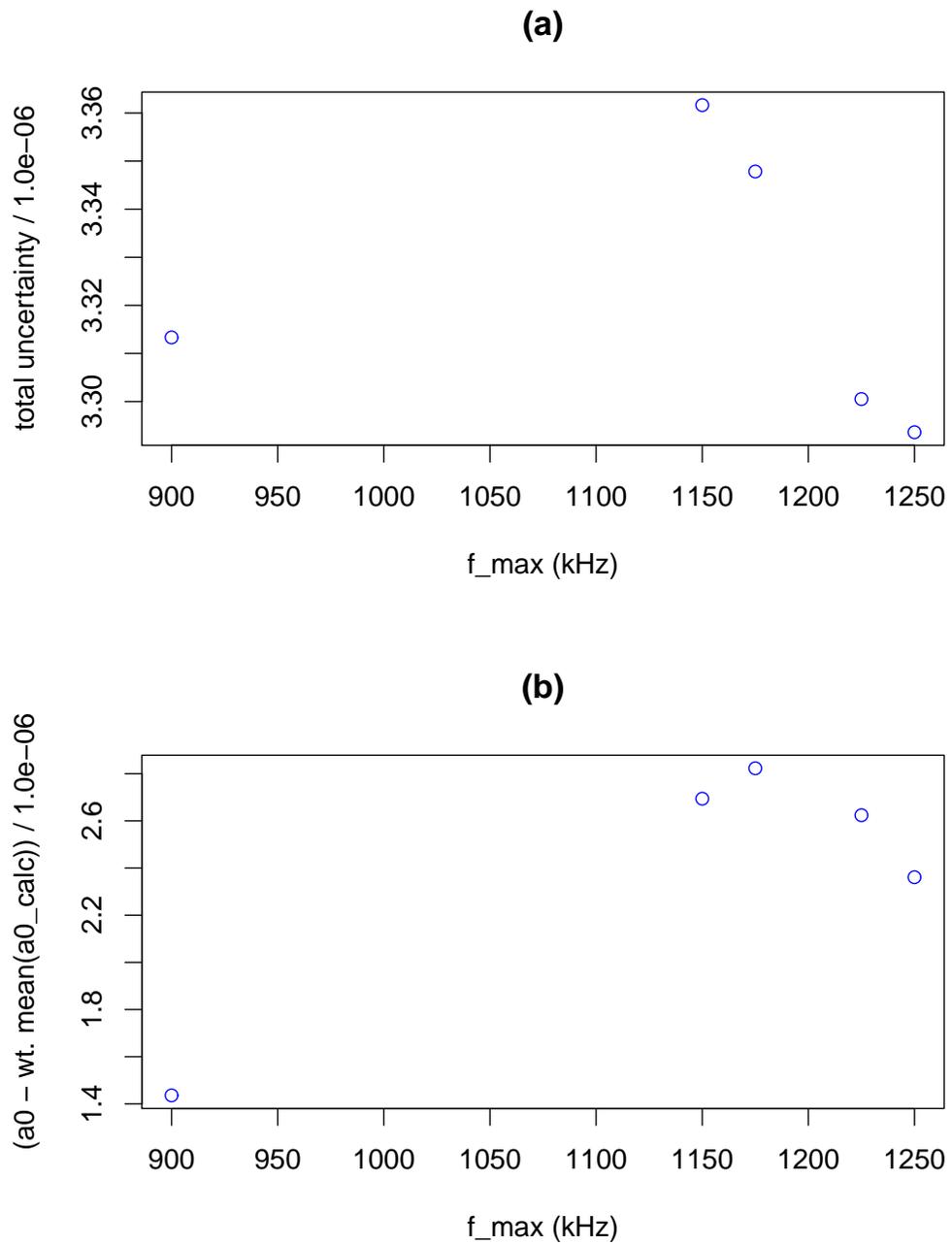}
\caption{Experimental data.
Results for $f_{\maxx}$ values which yield the five lowest
values of $\hat{\sigma}_{\tott}$.
(a) $\hat{\sigma}_{\tott}$ versus $f_{\maxx}$.
(b) Estimated ${a}_{0} - {\bar{a}}_{0,\calcc}$ versus $f_{\maxx}$.
}
\label{fig3}
\end{figure}

\begin{figure}[!t]
\centering
\includegraphics[height=7.5in]{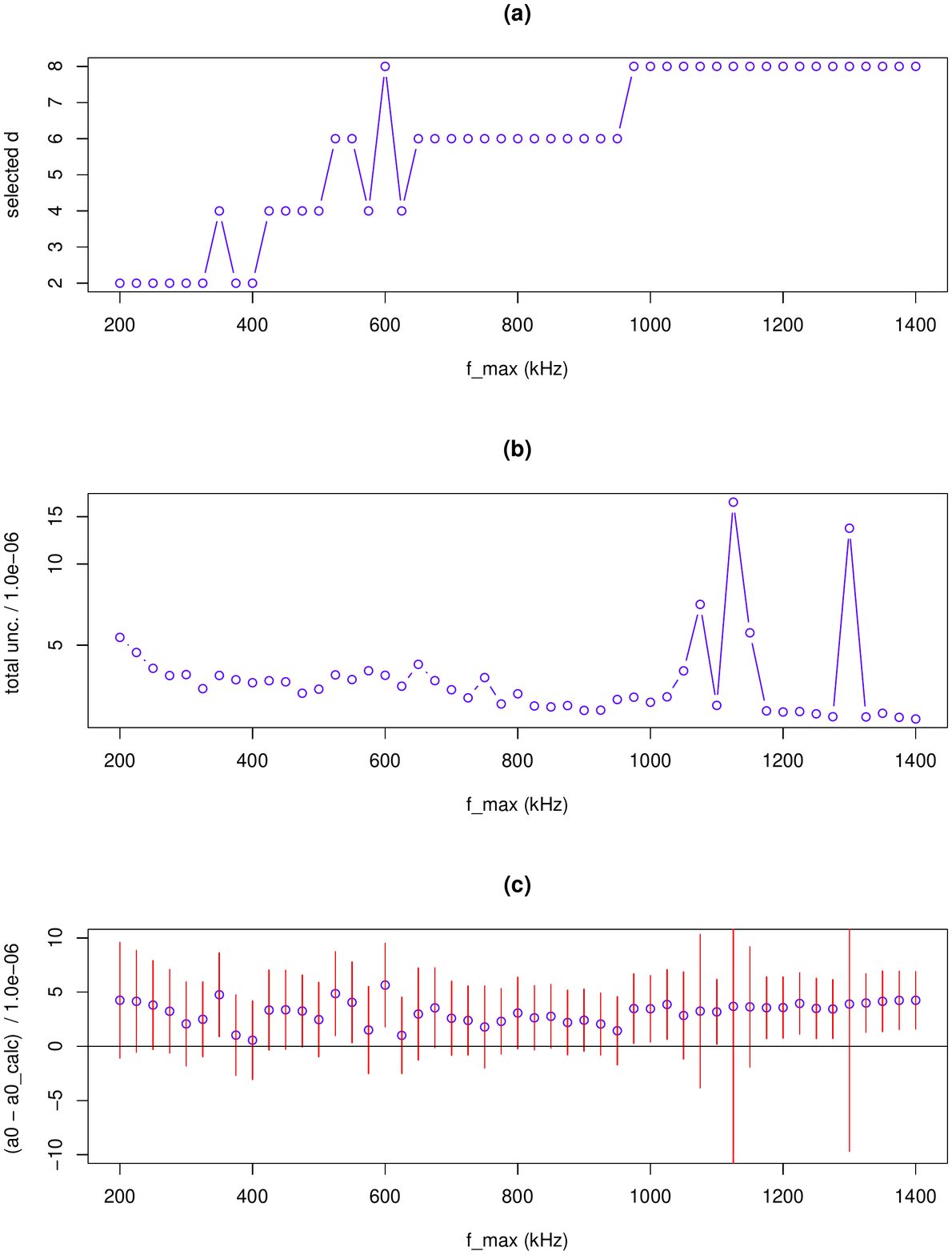}
\caption{Simulated data (realization 1). (a) Estimated polynomial complexity parameter $d$.
(b) Estimated total uncertainty $\hat{\sigma}_{\tott}$.
(c) Estimated $a_0 - a_{0,\calcc}$ and approximate 68 $\%$ coverage interval.
The true value
of
${a}_{0} - a_{0,\calcc}$ is 0.
}
\label{fig4}
\end{figure}

\begin{figure}[!t]
\centering
\includegraphics[height=7.5in]{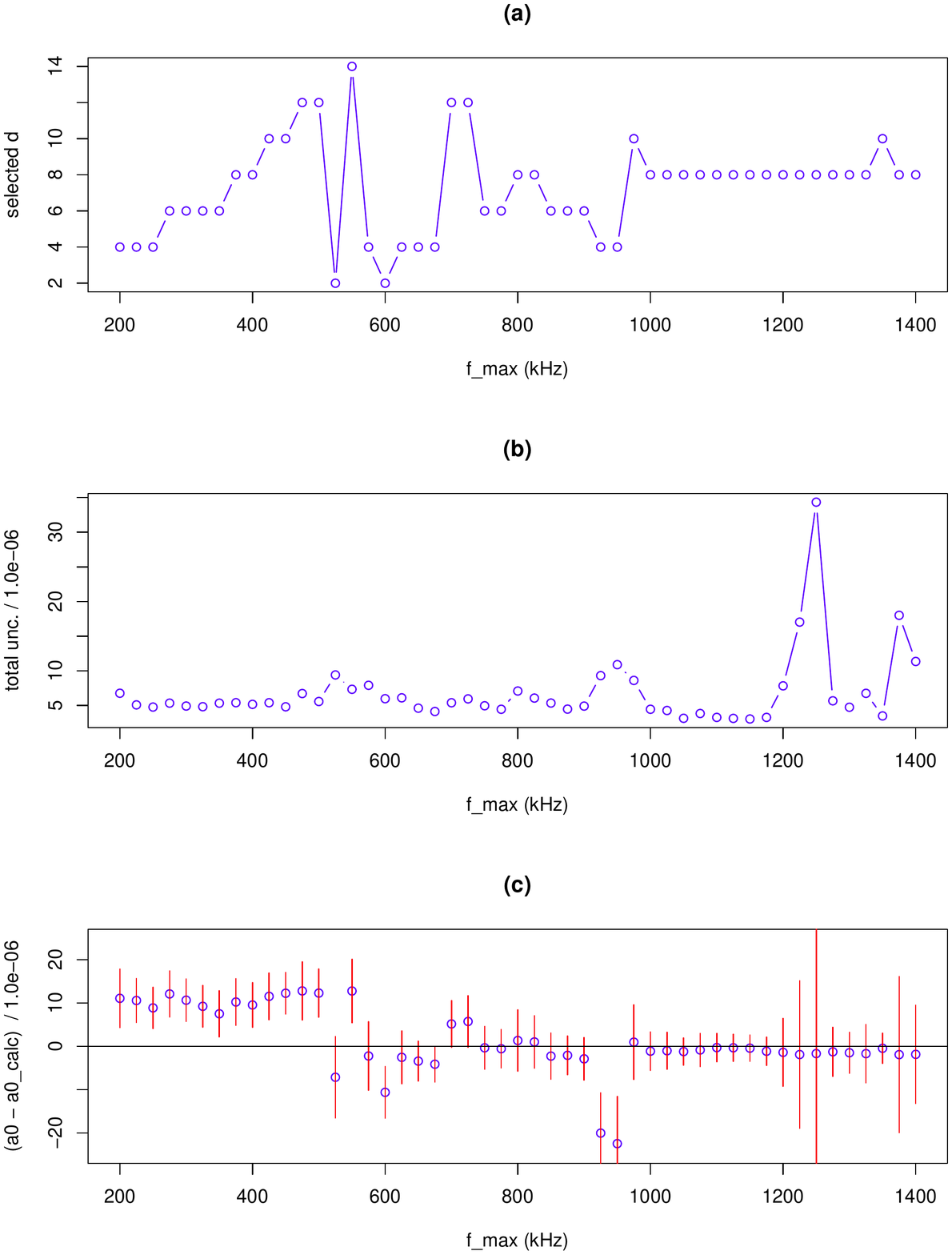}
\caption{Simulated data (realization 2). (a) Estimated polynomial complexity parameter $d$.
(b) Estimated total uncertainty $\hat{\sigma}_{\tott}$. 
(c) Estimated $a_0 - a_{0,\calcc}$ and approximate 68 $\%$ coverage interval.
The true value 
of
${a}_{0} - a_{0,\calcc}$ is 0.
}
\label{fig5}
\end{figure}

\begin{figure}[!t]
\centering
\includegraphics[height=7.5in]{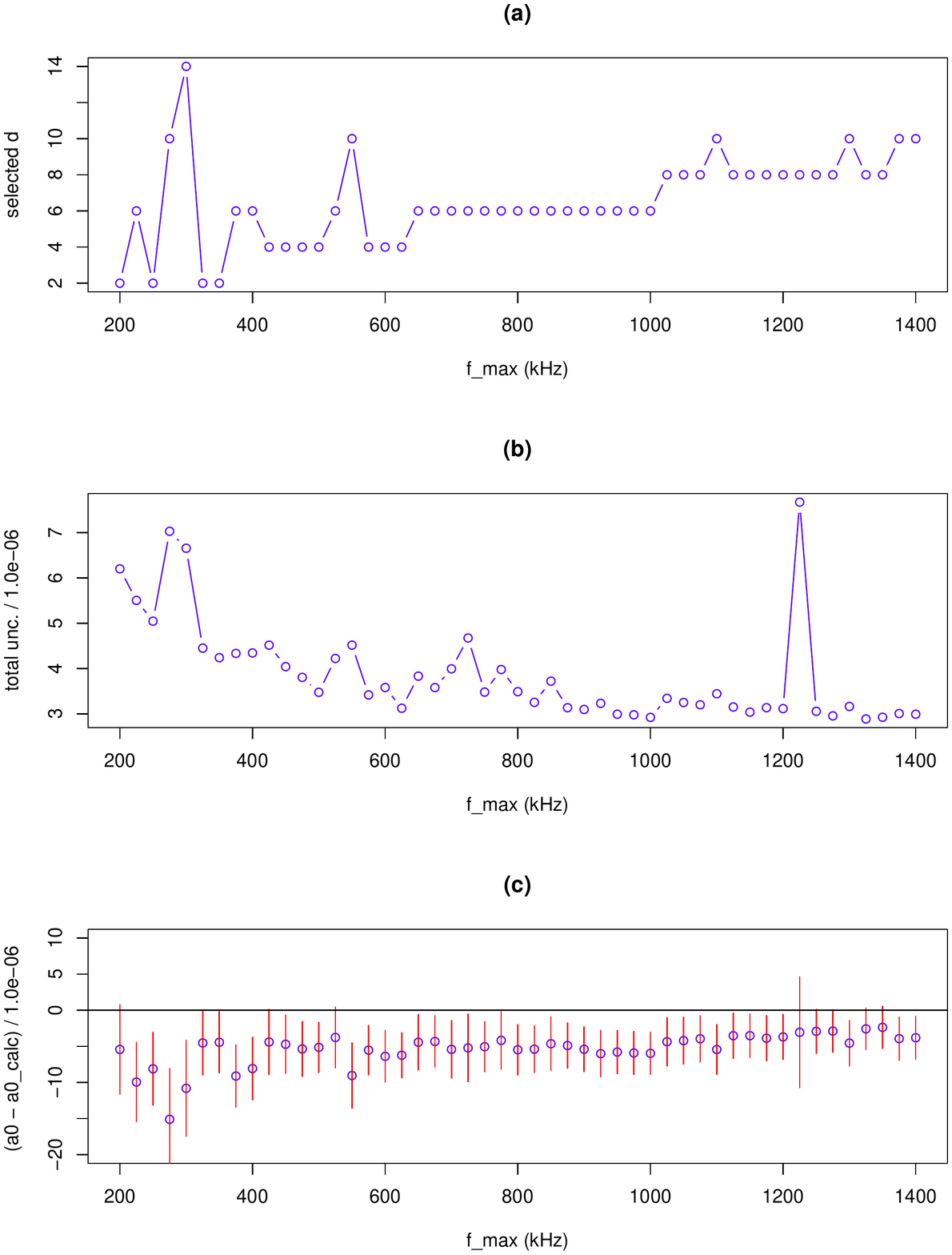}
\caption{Simulated data (realization 3). (a) Estimated polynomial complexity parameter $d$.
(b) Estimated total uncertainty $\hat{\sigma}_{\tott}$.
(c) Estimated $a_0 - a_{0,\calcc}$ and approximate 68 $\%$ coverage interval.
The true value 
of
${a}_{0} - a_{0,\calcc}$ is 0.
}
\label{fig6}
\end{figure}

\clearpage{}
\begin{figure}[!t]
\centering
\includegraphics[height=7.5in]{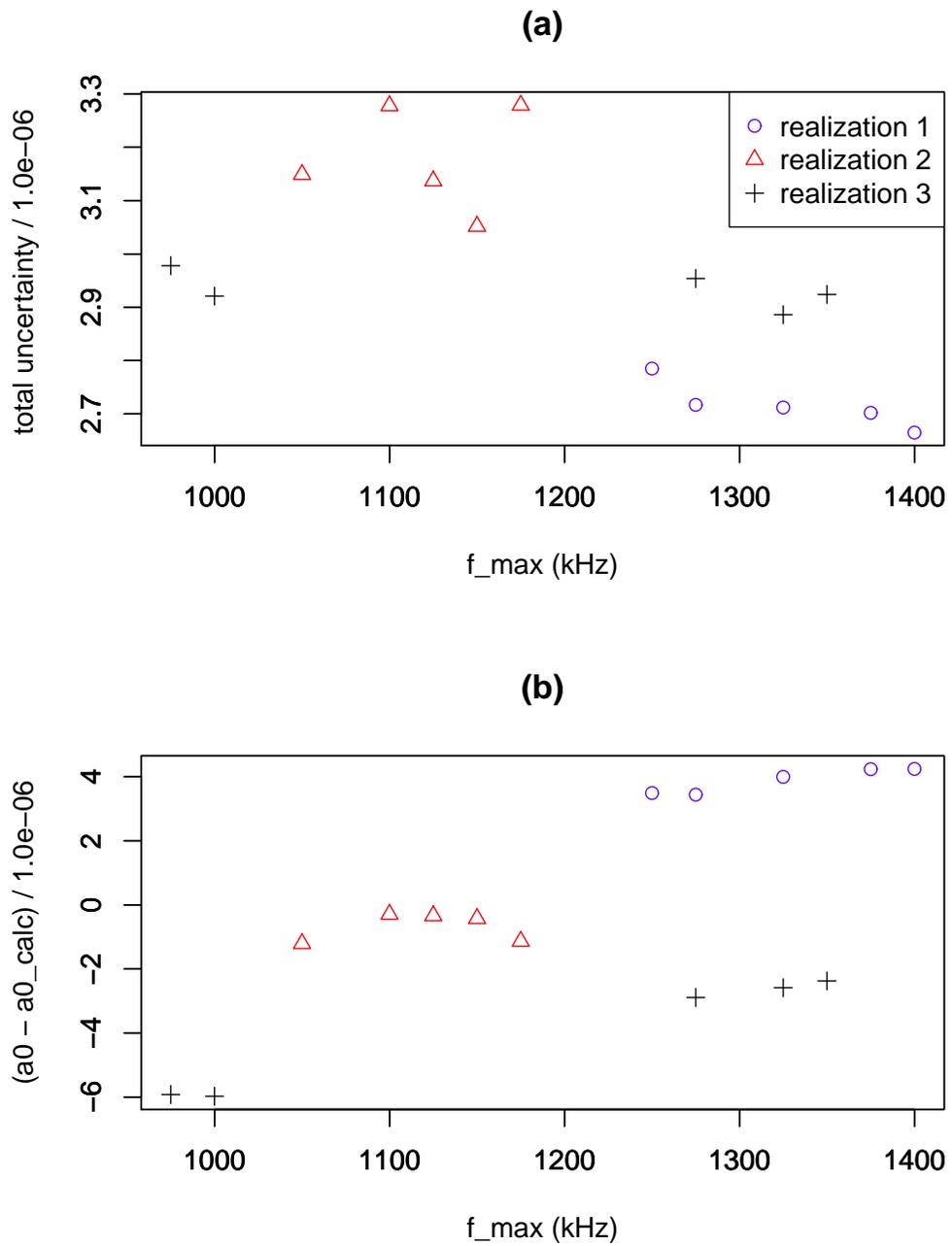}
\caption{Simulated data.
Results for $f_{\maxx}$ values which yield the five lowest
values of $\hat{\sigma}_{\tott}$.
(a) $\hat{\sigma}_{\tott} $ versus $f_{\maxx}$.
(b) estimated ${a}_{0} - {a}_{0,\calcc}$ versus $f_{\maxx}$.
The true value of
${a}_{0} - a_{0,\calcc}$ is 0.
}
\label{fig7}
\end{figure}

\clearpage{}
\begin{figure}[!t]
\centering
\includegraphics[height=7.5in]{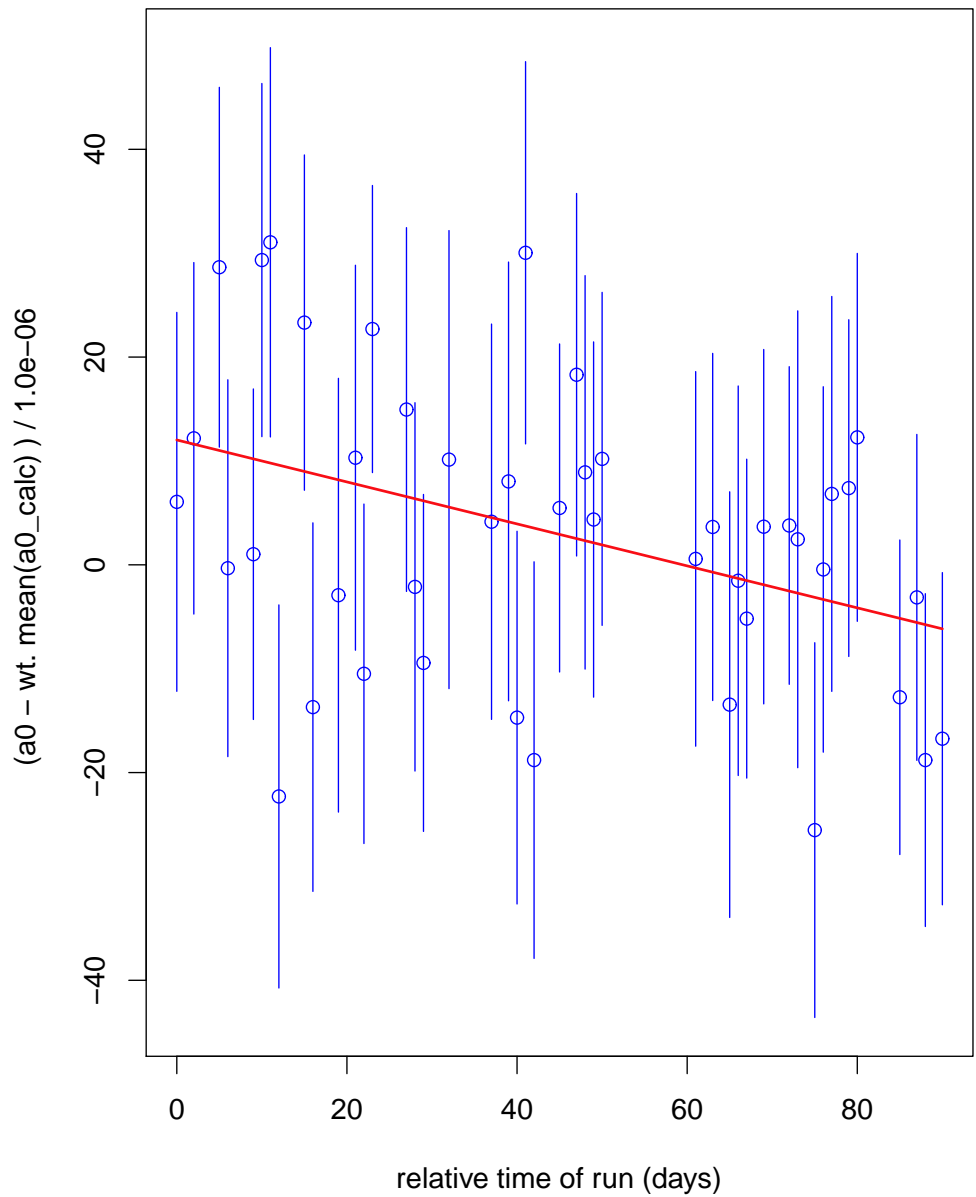}
\caption{From
the corrected ratio spectrum (Eq. 7), we estimate
${a}_{0} - \bar{a}_{0,\calcc}$  for each run
by fitting a $d=8$ ratio spectrum model
by the method of Least Squares.
The half-widths of the intervals
are asymptotic uncertainties determined by the Least Squares
method. The fitting bandwidth is
$f_{\maxx}$ = 1250 kHz.}
\label{fig8}
\end{figure}

\end{document}